\documentstyle[12pt]{article}
\input{epsf}
\newcommand{\be}{\begin{eqnarray}}
\newcommand{\ee}{\end{eqnarray}}
\newcommand{\ba}{\begin{array}}
\newcommand{\ea}{\end{array}}

\newcommand{\nablaboth}{\stackrel{\leftrightarrow}{\nabla}}
\begin{document}
\renewcommand{\thefootnote}{\fnsymbol{footnote}}
%
%

\rightline{RUB-TP2-14/02}
\vspace{0.5cm}
\begin{center}
{\Large Generalized parton distributions and strong forces inside nucleons and nuclei}\\
\vspace{0.35cm}
 M.V. Polyakov$^{a,b}$\\

\vspace{0.35cm}
$^a$Petersburg Nuclear Physics
Institute, Gatchina, St.\ Petersburg 188350, Russia\\
and\\
$^b$Institut f\"ur Theoretische Physik II,
Ruhr--Universit\"at Bochum, D--44780 Bochum, Germany

%
%
\end{center}

\begin{abstract}
We argue that generalized parton distributions (GPDs), accessible
in hard exclusive processes, carry information about the spatial
distribution of forces experienced by quarks and gluons inside
hadrons. This way the measurements of hard exclusive processes
open a possibility for direct ``measurements" of strong
forces in different parts of nucleons and nuclei. Also such studies
open a venue for addressing questions of the properties of the
quark (gluon) matter inside hadrons and nuclei. We give a simple
example of relations between GPDs and properties of ``nuclear
matter"
in finite nuclei.

\end{abstract}
\vspace{0.1cm}

\noindent
{\bf 1.}~The generalized parton distributions (GPDs), accessible in hard exclusive
reactions (see the original works \cite{DM,Ji,Col97,Rad97} and reviews
\cite{Jirev,Radrev,GPV,Belitsky:2001ns}), describe the response of
the target hadron to the well-defined QCD operators on the light-cone.
Generically the GPDs contain information about the matrix elements
of the following type:

\be
\label{AtoB}
\langle B| \bar \psi_\alpha(0)\ {\rm P}e^{ig\int_0^z dx_\mu A^\mu}\
\psi_\beta (z) |A\rangle\, ,\qquad
\langle B| G_{\alpha \beta}^a(0)\ \Biggl[{\rm P}e^{ig\int_0^z dx_\mu A^\mu}\Biggr]^{ab}\
G_{\mu \nu}^{b}(z) |A\rangle\, ,
\ee
where the operators are on the light-cone, i.e.
$z^2=0$, and $A,B$ are various hadronic states. In this way the hard exclusive processes provide us
with a set of new fundamental probes of the hadronic structure.
An important question is a physical interpretation of these probes. Many
ideas have already been put forward, for example, viewed in the
infinite momentum frame GPDs allow us to
probe the distribution of partons in the transverse plane this
way we can obtain detailed spatial partonic images
of hadrons, see e.g. \cite{Bur,Ral,Die}.
In this note we shall discuss why the lowest Mellin moments of
GPDs provide us with information about the spatial
distribution of energy, momentum and forces experienced by quarks and gluons inside
hadrons.

To be specific we consider a spin-$1/2$ hadronic target,
e.g. a nucleon. For the GPDs we shall use the notation of X.~Ji, see
Ref.~\cite{Jirev}. We note that all spin independent equations in this paper
apply to the spin-$0$ targets as well.

\vspace{0.1cm}
\noindent
{\bf 2.}~The $x$-moments of the GPDs play a special role as
they are related to the form factors of the symmetric energy
momentum tensor. The nucleon matrix element
of the
symmetric energy momentum tensor is characterized by three
scalar form factors \cite{Pagels,Ji}. The nucleon matrix elements
of the quark and gluon parts of the QCD energy-momentum tensor (EMT) can be
parametrized the following way \cite{Ji}:

\be
\nonumber
\langle p'| \hat T_{\mu\nu}^{Q,G}(0)|p\rangle&=&\bar N(p')\biggl[
M_2^{Q,G}(t)\ \frac{\bar P_\mu\bar P_\nu}{m_N}+
J^{Q,G}(t)\ \frac{i\bar P_{\{\mu}\sigma_{\nu\} \rho}\Delta^\rho}{m_N}
\\
&+&d^{Q,G}(t)\  \frac{1}{5 m_N}\ \left( \Delta_\mu\Delta_\nu-g_{\mu\nu}\Delta^2\right)
\pm \bar c(t)g_{\mu\nu}
\biggr]N(p)\, .
\label{EMTffs}
\ee
Here $\hat T_{\mu\nu}^Q=\frac i2\ \bar \psi \gamma_{\{\mu}
\nablaboth_{\nu\}}\psi$ is the quark part of the QCD
energy-momentum tensor (massless case) and $\hat T_{\mu\nu}^G=
G_{\mu\alpha}^aG_{\alpha\nu}^a+\frac 14 g_{\mu\nu}G^2$ is the
gluon part of the  QCD EMT. Dirac spinors $\bar N$ and $N$ are normalized by
$\bar N N=2\ m_N$ and the kinematical variables are defined as $\bar P=(p+p')/2$, $\Delta=(p'-p)$,
$t=\Delta^2$.  The form factor $\bar c(t)$ accounts for nonconservation of the separate
quark and gluon parts of the EMT. This form factor enters the quark and
gluon parts with opposite signs in order
to account for conservation of the total (quark+gluon) EMT.
The form factors in eq.~(\ref{EMTffs}) are related to the Mellin
moments of GPDs through\footnote{We write explicitly results only for quarks, corresponding
expressions for the gluon GPDs are similar } \cite{Ji}:

\be
\label{jisr} \int_{-1}^1 dx\ x\ \left(
H(x,\xi,t)+E(x,\xi,t)\right)=2J^Q(t)\, ,\\
\int_{-1}^1 dx\ x\
H(x,\xi,t)=M_2^Q(t) +\frac 45\ d^Q(t)\ \xi^2\, . \nonumber \ee
Such relations between GPDs and the form factors of EMT open
a possibility to study these form factors in hard exclusive processes.
For example, the form factor $d^Q(t)$ contributes to the $x_{Bj}$ independent part of the
real part of the DVCS amplitude, which is accessible
through the beam charge asymmetry \cite{Brodsky}. Simultaneously this form factor corresponds to
the first coefficient in the Gegenbauer expansion of the so-called D-term \cite{PW99} in the
parametrizations of the GPDs, see for details \cite{PW99,Kiv}.
The real part of the DVCS amplitude (spin-0 target for
simplicity) at small $x_{Bj}$ and $t$ to the leading order in
$\alpha_s(Q)$ can be written,
under certain simplifying assumptions, as:

\be
{\rm Re}A\propto \pi\ H(\xi,\xi,t)
\tan\left(\frac{\pi\omega}{2}\right)+2\ d^Q(t)+\ldots\, ,
\label{amplutide}
\ee
where $\omega$ corresponds to the exponent in the small $x_{Bj}$ behaviour of the singlet quark distribution
$q(x)+\bar q(x)\sim 1/x^{1+\omega}$. The ``slice" $H(\xi,\xi,t)$ of quark GPD is directly measurable
in the DVCS beam spin asymmetry. Ellipsis in eq.~(\ref{amplutide})
stands for the higher Gegenbauer coefficients of the
D-term expansion, which die out logarithmically with increasing of the photon virtuality and seems
to be small even at a low normalization point, see estimates in ref.~\cite{Kiv}.
Note that the constant contribution $\sim d^Q(t)+\ldots$ is
similar to the contribution of fixed poles in the angular momentum plane
to the virtual Compton scattering discussed, e.g. in
\cite{Cheng:vg,Brodsky}.

Let us now analyze the physics content of the form factors $M_2^Q(t)$, $J^Q(t)$ and $d^Q(t)$.
To reveal the physics content of these form factors, in the same way as for electromagnetic form factors \cite{Sachs},
it useful to consider the nucleon matrix elements of the energy momentum tensor in Breit frame.
In this frame the energy transfer $\Delta^0=0$, therefore one can introduce the static energy momentum tensor defined
as:

\be
T_{\mu\nu}^Q(\vec r,\vec s)=\frac{1}{2 E}\int \frac{d^3\Delta}{(2\pi)^3}\
e^{i\vec{r}\cdot\vec{\Delta}}\
\langle p',S'| \hat T_{\mu\nu}^Q(0)|p,S\rangle\, ,
\ee
where $\hat T_{\mu\nu}^Q(0)$ is the QCD operator of the symmetric energy
momentum tensor of quarks. In Breit frame $E=E'=\sqrt{m_N^2+\vec \Delta^2/4}$. The
polarization vectors of the initial and final nucleons, $S^\mu$ and $S'^\mu$
we choose in such a way that
both of them correspond to the same polarization vector $(0,\vec s)$ in the rest frame of the corresponding nucleon.
We also note that in the Breit frame the four-momentum transfer squared $t=-\vec \Delta^2$.

Various components of the static energy momentum tensor $T_{\mu\nu}^Q(\vec r,\vec s)$
can be interpreted as spatial distributions (averaged over time)
of the quark contribution to various mechanical characteristics of the nucleon.
However, doing this we have to be careful because of nonconservation of the separate quark and gluon
parts of the EMT encoded in the form factor $\bar c(t)$ in eq.~(\ref{EMTffs}). The point is that
the ``charges" obtained from the tensor densities through the relation like

\be
Q_\mu\left(x^0\right)=\int d^3r\ D_{0 \mu}\left(x^0,\vec
r\right)\,.
\ee
are time dependent for the non-conserved tensor density $D_{\nu\mu}$ and therefore the Lorentz covariance
generically is broken. Nevertheless for the
case of the quark or gluon part of the EMT one is free from
such a problem for the $T_{0 k}^Q(\vec r,\vec s)$ and
$\left(T_{ik}^Q(\vec r,\vec s)-\frac 13 \delta_{ik}T^Q_{ll}\right)$
components of the static tensor densities, because the ``problematic" term $\bar c(t)$ drops out
in these combinations.

The components
$T_{0 k}^Q(\vec r,\vec s)$ correspond to the distribution of the quark momentum in the nucleon.
The components $\left(T_{ik}^Q(\vec r,\vec s)-\frac 13 \delta_{ik}T^Q_{ll}\right)$ characterize the spatial distribution (averaged over time)
of ``shear forces" experienced by quarks inside the nucleon. For a spin-1/2
hadron,
only the component $T_{0 k}^Q(\vec r,\vec s)$ is sensitive to the polarization state.
For the higher spin hadrons (e.g. higher spin nuclei) all components of $
T_{\mu\nu}^Q(\vec r,\vec s,\ldots)$ are polarization dependent.

Now we can easily relate the form factors $J^Q(t)$ and $d^Q(t)$ to the spatial distribution of
the energy-momentum and forces encoded in $T_{\mu\nu}^Q(\vec r,\vec s)$. The relation
for the form factor $J^Q(t)$ is the following:

\be
\label{enmom}
J^Q(t)+\frac 23\ t\ \frac{dJ^Q(t)}{dt}= \int d^3 r\ e^{-i\vec{r}\cdot\vec{\Delta}}\ \varepsilon^{i j k}\ s_i\ r_j\ T_{0k}^Q(\vec r,\vec s)\, .
\ee
We see that the form factor $J^Q(t)$ gives us information
about the spatial distribution
of the quark angular momentum\footnote{Note that
$\varepsilon^{i j k}\ r_j\ T_{0k}^Q(\vec r,\vec s)$ corresponds to angular momentum density.} inside the nucleon.
Now if we take the limit  $t\to 0$ in eq.~(\ref{enmom}) we obtain:

\be
\label{normal}
J^Q(0)&=& \int d^3 r\ \varepsilon^{i j k}\ s_i\ r_j\ T_{0k}^Q(\vec r,\vec s)\, .
\ee
This relation illustrates the interpretation of $J^Q(0)$
as a fraction of the angular momentum of the
nucleon carried by quarks and antiquarks \cite{Ji}.

Concerning the form factor $M_2^Q(t)$, one can easily see, going to the infinite momentum
frame in eq.~(\ref{EMTffs}), that at $t=0$ it is related to the momentum fraction carried by
quarks which is measured in inclusive deep inelastic scattering. The constant $M_2^Q(0)$
is related to the parton distributions via
$
M_2^Q(0)= \sum_q \int_0^1dx\ x \ \left(q(x) + \bar q(x) \right)\, .
$

Obviously the form factors
$M_2^Q(t)$, $J^Q(t)$ and $d^Q(t)$ are renormalization scale dependent.
This corresponds to the fact that
the individual distributions of quarks (gluons) depend on the resolution scale.
The scale independent
quantities are obtained adding the contributions of quarks and gluons. These are $M_2(t)=M_2^Q(t)+M_2^G(t)$,
$J(t)=J^Q(t)+J^G(t)$ and $d(t)=d^Q(t)+d^G(t)$. The  scale independent form factors $M_2(t)$ and $J(t)$
are expressible in terms of the total static energy momentum tensor $T_{\mu\nu}(\vec r,\vec s)=
T_{\mu\nu}^Q(\vec r,\vec s)+T_{\mu\nu}^G(\vec r,\vec s)$\footnote{Note that for the total
conserved EMT all components of this tensor have a meaning of ``good" tensor densities}.
The corresponding expressions
have the form
\be
\nonumber
M_2(t)-\frac{t}{4m_N^2}\left[
M_2(t)-2 J(t)+\frac 45\ d(t)\right]=\frac{1}{m_N}\
\int d^3 r\ e^{-i\vec{r}\cdot\vec{\Delta}}\ T_{00}(\vec r,\vec s)\, ,\\
J(t)+\frac 23\ t\ \frac{dJ(t)}{dt}= \int d^3 r\ e^{-i\vec{r}\cdot\vec{\Delta}}\ \varepsilon^{i j k}\ s_i\ r_j\
T_{0k}(\vec r,\vec s)\, ,
\label{EMTMJ}
\ee
which at $t=0$ read

\be
\nonumber
M_2(0)&=&\frac{1}{m_N}\ \int d^3 r\ T_{00}(\vec r,\vec s)=1\\
\nonumber
J(0)&=& \int d^3 r\ \varepsilon^{i j k}\ s_i\ r_j\ T_{0k}(\vec r,\vec s)=\frac 12\, .
\ee
Written in this form the above equations have obvious interpretation.
The first one tells us that the total energy of
the nucleon in the rest frame is equal to its mass. The second equation states that the total spin
of the nucleon is 1/2. Also it shows that the anomalous gravimagnetic moment
of the nucleon is zero
\cite{Okun}.
\vspace{0.1cm}

\noindent
{\bf 3.}~Now let us turn to the physics content of the form factor $d^Q(t)$. It is easy to see that this form factor
is related to the traceless part of the static stress tensor $T_{ik}^Q(\vec r,\vec s)$ which
characterizes the spatial distribution
(averaged over time) of shear forces experienced by quarks in the nucleon \cite{izumrud}.
In detail this relation is
the following:

\be
\label{EMTd1}
d^Q(t)+\frac 4 3\ t \frac{d }{dt}d^Q(t)
+\frac{4}{15}\ t^2 \frac{d^2 }{dt^2}d^Q(t)
=-\frac{m_N}{2}\  \int d^3r\ e^{-i\vec{r}\cdot\vec{\Delta}}\  T_{ij}^Q(\vec r)\ \left(r^i r^j-\frac
13\
\delta^{ij} r^2\right)\, .
\ee
If one considered the nucleon as a continuous medium then $T_{ij}^Q(\vec r)$
would characterize the force experienced by quarks in an infinitesimal volume
at distance $\vec r$ from the centre of the nucleon.
At $t=0$  eq.~(\ref{EMTd1}) gives:
\be
\label{EMTd0}
d^Q(0)
=-\frac{m_N}{2}\  \int d^3r\   T_{ij}^Q(\vec r)\ \left(r^i r^j-\frac
13\
\delta^{ij} r^2\right)\, .
\ee
The expressions for the gluon and total energy momentum tensors are analogous .

As our understanding of the forces inside hadrons in QCD
is still rather limited we can not make first principles prediction
for the value of $d(t)$. The estimate
which is based on the calculation of GPDs
in the chiral quark soliton model \cite{Pet98}
at a low normalization point $\mu\approx 0.6$~GeV,
gives \cite{Kiv,Peterw} a rather large and negative value of $d^Q(0) \approx
-4.0$. The negative value of this
constant has a deep relation to the spontaneous breaking of chiral symmetry
in QCD, see \cite{MVP98,Kiv,GPV}.
\vspace{0.1cm}

\noindent
{\bf 4.}~To illustrate physics behind the form factor
$d(t)=d^Q(t)+d^G(t)$ let us consider an idealized model of a very large nucleus.
Generically the static stress tensor for spin-0 and spin-1/2 targets can be decomposed as:
\be
\label{tijdecomp}
T_{ij}(\vec r)=s(r) \left(\frac{r_ir_j}{r^2}-\frac 13\ \delta_{ij}\right)+p(r)\delta_{ij}\, .
\ee
The functions $s(r)$ and $p(r)$ are related to each other by conservation of the
total energy-momentum tensor. The function $p(r) $ can be
interpreted as the radial distribution of the ``pressure" inside the hadron\footnote{Note that
for non-homogeneous media the pressure, defined as the force per unit area and
directed orthogonally
to the surface element, gets a contribution from shear forces as well.
}.
The function $s(r)$ is related to the distribution of the shear forces
and, in the simple model of a large nucleus considered below,
is related to the surface tension.

For a very large nucleus  we can assume that the pressure $p(r)$
is
constant, $p_0$, in the bulk of the nucleus, and it changes only in the thin ``skin"
around radius $R$ of the nucleus. It is known from the famous electron scattering experiments
\cite{Hofstadter}
that the distribution of electric charge has such a shape in large nuclei.
Surely this does not imply
that, say, the distribution of pressure follows the shape of the electric charge
distribution, although such an assumption
is rather sensible. The measurements of coherent hard exclusive processes (like DVCS) on nuclei will provide
us with detailed information about deviations of the energy, pressure, and shear forces distributions
from that of electric charge, see eqs.~(\ref{EMTMJ},\ref{EMTd1}).
Since here our aim is merely illustrative,
we consider an (over)idealized case of a nucleus with sharp edges, like a liquid drop.
In this case, the pressure can be written
as\footnote{One can easily include higher terms which take into account the width of the ``skin"
and other characteristics of the spatial distribution of forces, see below.}:

\be
p(r)=p_0\ \theta(R-r)-\frac{p_0 R}{3} \delta(R-r)\, .
\ee
{}From the condition $\partial_k T_{kl}(\vec r)=0$ we obtain that
\be
\label{sdel}
s(r)=\frac{p_0 R}{2} \delta(R-r)=\gamma\ \delta(R-r)\, .
\ee
This equation immediately shows that the function $s(r)$ in eq.~(\ref{tijdecomp}) has a meaning of the
surface tension $\gamma$ of the nucleus\footnote{Recall
the well known Kelvin relation between
the pressure in a liquid spherical drop, its surface tension and the radius of the drop $P=2\gamma/R$
\cite{Kelvin}.}.
Substituting the solution (\ref{sdel}) into eq.~(\ref{EMTd0})
we obtain for the constant
$d(0)$ the following value:

\be
\label{d0estimate}
d(0)=-\frac{4\pi}{3} m_A\ \gamma\ R^4\, .
\ee
First we see that the $d$-constant is negative.
The effect of the finite width of the nuclear ``skin"
also has a negative sign. The corresponding formula can be easily derived:

\be
\label{d0estimateWidth} d(0)=-\frac{4\pi}{3} m_A\ \gamma\ R^4\
\left(1+\frac{5\pi^2}{3}\ \frac{a^2}{R^2}\right)\, , \ee where $a$
is a ``skin" width introduced by replacing the step function by
Fermi-like function $\theta(R-r)\to 1/(1+\exp((r-R)/a))$. If we
assume that the surface tension depends slowly on the atomic
number (as it is suggested by the Weizs\"acker formula), we come
to the conclusion that $d(0)\sim A^{7/3}$, i.e. it rapidly grows
with the atomic number. This fact implies that the contribution of
the D-term to the real part of the DVCS amplitude grows with an
increase of the atomic number as $A^{4/3}$. This should be
compared to the behaviour of the amplitude $\sim A$ in the impulse
approximation. If true, rather interesting phenomenon! In
principle, it can be checked by measuring the charge beam
asymmetry in coherent DVCS on nuclear targets. Taking the value of
the nuclear surface tension of $\gamma\approx 1 MeV/fm^2$ (as it follows
from the Weizs\"acker formula) and Hofstadter's $R=1.12\ A^{1/3}
fm$, $a\approx 0.54 fm$ \cite{Hofstadter}, we get an estimate
$d(0)\approx -0.2\ A^{7/3}\left(1+3.8/A^{2/3}\right)$.
Although
being very rough and naive, estimate (\ref{d0estimateWidth})  shows
a big potential of hard exclusive processes for studies of properties
of quark and gluon ``matter" inside nuclei.

Another possible application of our
eqs.~(\ref{enmom},\ref{EMTMJ},\ref{EMTd1}) is the estimations of
the EMT form factors in various effective models of the nucleon structure, like chiral
soliton models. In the latter case the static EMT can be obtained
from the expression for EMT  of the effective chiral
Lagrangian (EChL) computed on the static soliton field.
\vspace{0.1cm}

\noindent
{\bf 5.}~Hard exclusive processes allow us to extend a set of fundamental probes of hadronic structure.
As an example we considered physics encoded in the lowest Mellin moments of generalized parton distributions.
In particular, we showed that one has an access to the spatial distributions of energy, angular momentum
and forces inside  hadrons, see eqs.~(\ref{enmom},\ref{EMTMJ},\ref{EMTd1}). These equations give us a tool
for systematic studies of the properties (distributions of energy, angular momentum, pressure, shear forces, etc.)
 of the quark-gluon ``matter" inside hadrons.

As an illustration, we considered a rough idealized picture of a large nucleus.
We showed that even in this picture the
parameters of GPDs carry detailed information about nuclear matter
in the nucleus-- knowledge which is still incomplete. We note that the picture
used here can be considerably refined, for instance one can, instead of
macroscopic approach used here, apply microscopic description.

Hopefully experimental
studies of the hard exclusive processes will fill the gap in our understanding
of the strong forces creating our world as we see it.
See the first experimental data on deeply virtual Compton
scattering (DVCS) \cite{ZEUS,H1,Amarian,JLAB,ChAsy}. We hope that this studies will be extended
for the nuclear targets.
\vspace{0.2cm}

{\it \small\it  I am grateful to M.~Amarian, L.~Frankfurt, K.~Goeke,
V.~Guzey, P.V. Pobylitsa
and M.~Strikman for valuable
discussions. The
work is supported
by the Sofja Kovalevskaja Programme of the Alexander von Humboldt
Foundation, the Federal Ministry of Education and Research and the
Programme for Investment in the Future of German Government.}

\end{document}